\documentclass[11pt,a4paper]{article}
\usepackage{geometry}
\usepackage[english]{babel}
\usepackage{amsmath,amsthm}
\usepackage{amsfonts}
\usepackage{graphicx}
\usepackage{bm}
\usepackage[colorlinks,linkcolor=blue,anchorcolor=blue,citecolor=blue,urlcolor=blue,filecolor=blue,menucolor=blue,runcolor=blue]{hyperref}
\usepackage{ulem}
\usepackage{authblk}

\begin{document}

\title{Erasable superconductivity in topological insulator Bi$_2$Se$_3$ induced by voltage pulse}

\author[1,2]{Tian Le \thanks{Corresponding author: tianlephy@iphy.ac.cn}}
\author[3]{Qikai Ye}
\author[2]{Chufan Chen}
\author[2]{Lichang Yin}
\author[2]{Dongting Zhang}
\author[3]{Xiaozhi Wang \thanks{Corresponding author: xw224@zju.edu.cn}}
\author[2,4,5]{Xin Lu \thanks{Corresponding author: xinluphy@zju.edu.cn}}

\affil[1] {\textit{Beijing National Laboratory for Condensed Matter Physics, Institute of Physics, Chinese Academy of Sciences, Beijing 100190, China}}
\affil[2] {\textit{Center for Correlated Matter and Department of Physics, Zhejiang University, Hangzhou 310058, China}}
\affil[3] {\textit{Key Labaratory of Advanced Micro/Nano Electronic Devices and Smart Systems of Zhejiang, College of Information Science and Electronic Engineering, Zhejiang University, Hangzhou 310027, China}}
\affil[4] {\textit{Zhejiang Province Key Laboratory of Quantum Technology and Device, Zhejiang University, Hangzhou 310027, China}}
\affil[5] {\textit{Collaborative Innovation Center of Advanced Microstructures, Nanjing University, Nanjing, 210093, China}}
\date{}

\maketitle

\textbf{Three-dimensional topological insulators (TIs) attract much attention due to its topologically protected Dirac surface states. Doping into TIs or their proximity with normal superconductors can promote the realization of topological superconductivity (SC) and Majorana fermions with potential applications in quantum computations. Here, an emergent superconductivity was observed in local mesoscopic point-contacts on the topological insulator Bi$_2$Se$_3$ by applying a voltage pulse through the contacts, evidenced by the Andreev reflection peak in the point-contact spectra and a visible resistance drop in the four-probe electrical resistance measurements. More intriguingly, the superconductivity can be erased with thermal cycles by warming up to high temperatures (300 K) and induced again by the voltage pulse at the base temperature (1.9 K), suggesting a significance for designing new types of quantum devices. Nematic behaviour is also observed in the superconducting state, similar to the case of Cu$_x$Bi$_2$Se$_3$ as topological superconductor candidates.}

So far, various methods such as doping, proximity effect, hydraulic pressure, tip-contact have been applied on nontrivial topological materials in order to induce topological superconductivity and Majorana fermions \cite{PhysRevLett.104.057001, wang2012coexistence, veldhorst2012josephson, wang2013fully, PhysRevLett.111.087001, aggarwal2016unconventional, wang2016observation}. Majorana fermions are proposed to play a crucial role in the fault-tolerant quantum computation \cite{RevModPhys.80.1083, alicea2011non, PhysRevLett.105.077001}. Among them, Bi$_2$Se$_3$ has served as a characteristic compound of topological insulators susceptible to tuning and emergence of superconductivity \cite{PhysRevLett.104.057001, wang2012coexistence, PhysRevLett.111.087001} and its van der Waals (vdW) structure also implies a potential application in quantum electronic devices, especially when superconductivity can be achieved \cite{geim2013van, liu2019van, basov2016polaritons}. However, an easy and controllable condition to realize SC in Bi$_2$Se$_3$ is still lacking and thus desirable.

In this paper, an unambiguous superconductivity is observed for the mesoscopic point-contacts on topological insulator Bi$_2$Se$_3$ after applying a voltage pulse, evidenced by the Andreev reflection peak in point-contact spectra (PCS) and a resistance drop for the exfoliated Bi$_2$Se$_3$ flake samples. Superconductivity emerges only in the case of Ag-Bi$_2$Se$_3$ contacts with either silver paints in soft-PCS(SPCS) or Ag tips in mechanical-PCS(MPCS) right after the voltage pulse, but is absent for the Au, Cu or Ti tips, favoring the scenario that the emergent SC is probably due to Ag dopants into the vdW gap of Bi$_2$Se$_3$ across the interface. The SC becomes unstable against thermal cycles and disappears when the sample is warmed up to high temperatures, however, it can be re-induced by a new voltage pulse after cooled back to 1.9 K. Our observations strongly support the erasable nature of the SC in the local contact region, ensuring potential applications in quantum devices.

\begin{figure}
\centerline{\includegraphics[angle=0,width=0.8\textwidth]{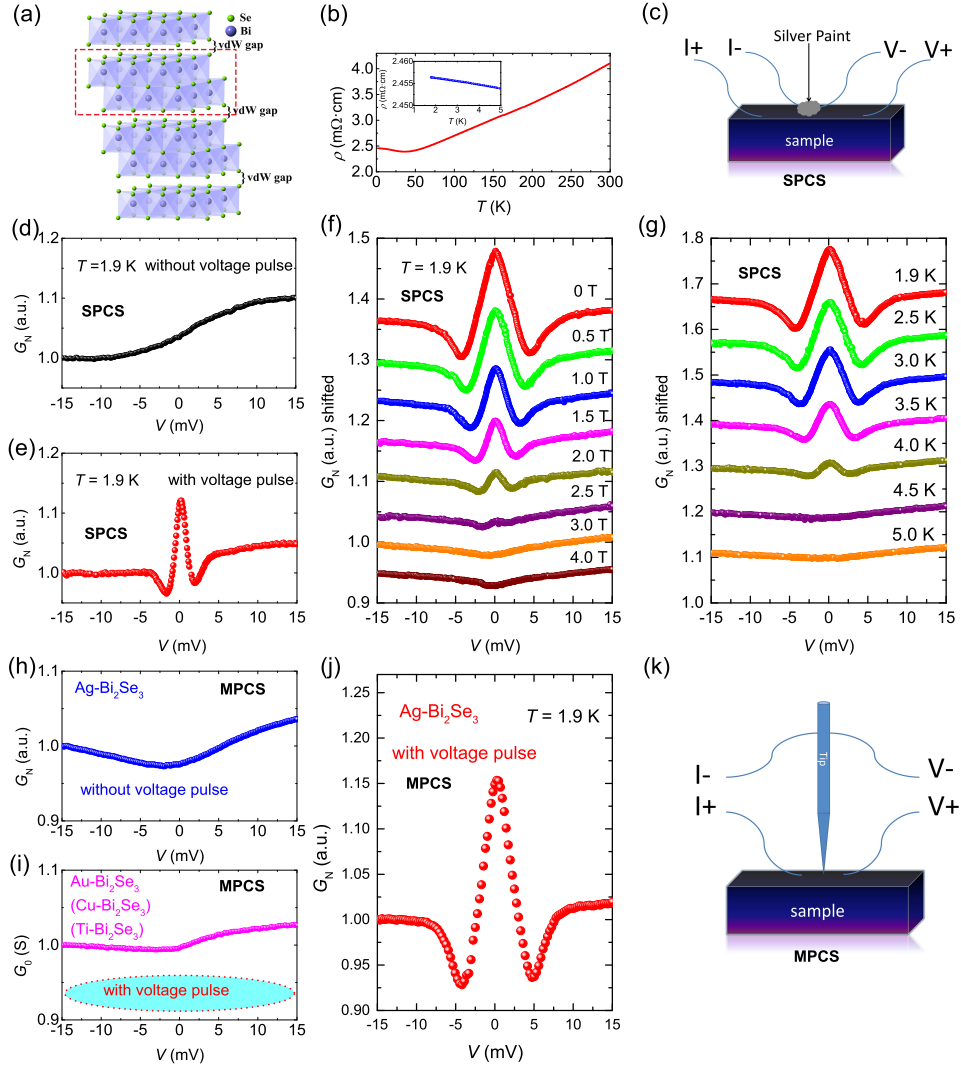}}
\vspace{-3pt} \caption{\label{Figure1}\textbf{Evidence for induced superconductivity in mesoscopic Ag point-contact on Bi$_2$Se$_3$.} \textbf{(a)} The crystal structure of topological insulator Bi$_2$Se$_3$ with vdW gap between adjacent Se-Bi-Se-Bi-Se quintuple layers. \textbf{(b)} Temperature dependence of the electrical resistivity for the pristine Bi$_2$Se$_3$ sample with a dimension of 1073 $\mu$m $\times$ 591 $\mu$m $\times$ 6.3 $\mu$m. \textbf{(c)} Schematic illustration of the soft-PCS electrode configuration on Bi$_2$Se$_3$. \textbf{(d)} A representative soft-PCS conductance curve for silver paint on the Bi$_2$Se$_3$ crystal at 1.9 K without any voltage pulse. $G\rm_N$ is the normalized conductance divided by the conductance value at high voltages. \textbf{(e)} A reproducible ZBCP feature in the soft-PCS at 1.9 K after voltage pulse on the contact. \textbf{(f) \& (g)} Magnetic field and temperature evolution of the soft-PCS conductance curves with induced superconductivity in the contact region, respectively. \textbf{(h) \& (i)} Mechanical-PCS conductance curve at 1.9 K for Ag tip before voltage pulse and for Au, Cu or Ti tip after voltage pulse, respectively.  \textbf{(j)} Mechanical-PCS conductance curve at 1.9 K for Ag tip with induced SC after voltage pulse. \textbf{(k)} Schematic illustration of the needle-anvil configuration of mechanical point-contacts. }
\vspace{-3pt}
\end{figure}

The Bi$_2$Se$_3$ crystal is composed of alternating Se-Bi-Se-Bi-Se quintuple layers as shown in Fig. 1a and there exists a weak vdW gap between adjacent quintuple layers, making this material easy to be exfoliated as other vdW materials. Temperature dependent resistivity of Bi$_2$Se$_3$ with a sample size of 1073 $\mu$m  $\times$ 591 $\mu$m  $\times$ 6.3 $\mu$m is shown in Fig. 1b, which has a slight upturn due to its insulating band gap but saturates below 40 K with its gapless surface state, consistent with previous reports \cite{PhysRevB.81.241301, PhysRevLett.109.166801}. Soft point-contacts are formed by silver paints on Bi$_2$Se$_3$ as in Fig. 1c, same as the soft-PCS configuration on Cu$_x$Bi$_2$Se$_3$ \cite{PhysRevLett.107.217001}. Fig. 1d shows the soft-PCS conductance curve on Bi$_2$Se$_3$ at 1.9 K and only an asymmetric background is observed in the absence of any special features. Interestingly, when a voltage pulse between 4 -10 V with a pulse duration less than 10 ms is applied across the contacts, a reproducible zero-bias conductance peak (ZBCP) can be observed with dips at high bias voltages as in Fig. 1e. Fig. 1f shows the evolution of conductance curves in magnetic field and the ZBCP is gradually suppressed without any peak splitting and finally disappears around 3.0 T. Moreover, the ZBCP gets weaker in intensity with increased temperatures untill around 4.5 K, unambiguously signaling the Andreev reflections and induced superconductivity in the contact area by voltage pulses \cite{PhysRevB.69.134507, aggarwal2016unconventional, wang2016observation}. The maximum superconducting transition temperature $T\rm_c$ for all contacts can be as high as 4.5 K as illustrated by the temperature dependent conductance curves on Bi$_2$Se$_3$ in Fig. 1g (or the temperature dependent zero-bias conductance (ZBC) in Fig. S1). We note here that the ZBCP only appears on the contact affected by the voltage pulse and other contacts without it on the same sample would not show any SC trace (Several soft point-contacts on the same sample can be realized in practice). The width of the ZBCP not only depends on the intensity of voltage pulse, but also influenced by detailed contact conditions, such as contact area and contact resistance, and more details can be referred in Supporting Information.

In comparison with our soft-PCS results, a needle-anvil type of mechanical-PCS was also applied as schematically illustrated in Fig. 1k with several tip options available such as silver, gold, copper or titanium. Similar behaviours are observed for the Ag tip on Bi$_2$Se$_3$ crystals after a voltage pulse, where a local SC in the contact can be induced with a characteristic SC ZBCP at 1.9 K. In sharp contrast, Au, Cu or Ti tips on Bi$_2$Se$_3$ fail to induce any SC within the maximum voltage pulse $\sim$ 10 V as in Fig. 1i. We would thus speculate that the induced SC in the point-contact area is intimately associated with Ag atoms as a result of voltage pulses.

In order to confirm the probable superconductivity with electrical resistive measurements, the Bi$_2$Se$_3$ sample was exfoliated by tape to a very tiny flake and it was transferred onto the pre-deposited Ag electrodes on the silicon substrate as shown in the inset of Fig. 2a, where the space between the neighboring electrode ends is about 3 $\mu$m in distance. The temperature dependent electrical resistance for the pristine device is shown in Fig. 2a, which also has an upturn at low temperatures as in Fig. 1b for the bulk sample. Once voltage pulse is applied on the Ag electrodes, a small drop of resistance shows up below 4.5 K ($\sim 1 -2\%$) as in Fig. 2a, implying an induced SC as in the PCS measurements. However, the resistance does not go to zero and it is natural to assume a local SC is induced only in the contact region for the Bi$_2$Se$_3$ sample. In Fig. 2b, the $T\rm_c$ decreases with increased magnetic field, and the SC transition is totally gone above 2.5 T.

It is interesting to study the stability of this local superconductivity against thermal cycles and the sample resistance over several thermal cycles with a cooling (warming) speed of 3 K/min is shown in Fig. 2c. For the first cycle, the sample is only warmed up to 50 K and then cooled back to 1.9 K, where the SC transition temperature $T\rm_c$ nearly maintains the same value of 4.5 K with the resistance curves overlapping on each other. However, as long as the sample is warmed up to a higher terminal temperature, $T\rm_c$ is gradually reduced and the transition finally disappears after warming up to 300 K. Moreover, superconductivity would emerge once again at 1.9 K after a new voltage pulse on Bi$_2$Se$_3$. Meanwhile, our soft-PCS with silver paint shows a consistent behaviour: the ZBCP would disappear in the conductance curves while the zero-bias conductance as a function of temperature shows the absence of any SC transition after thermal cycles as in Fig. 2d. It is thus a strong evidence for the metastable nature of induced SC in the mesoscopic Ag point-contacts on Bi$_2$Se$_3$, which can be repetitively erased by thermal activations.

\begin{figure}
\centerline{\includegraphics[angle=0,width=0.8\textwidth]{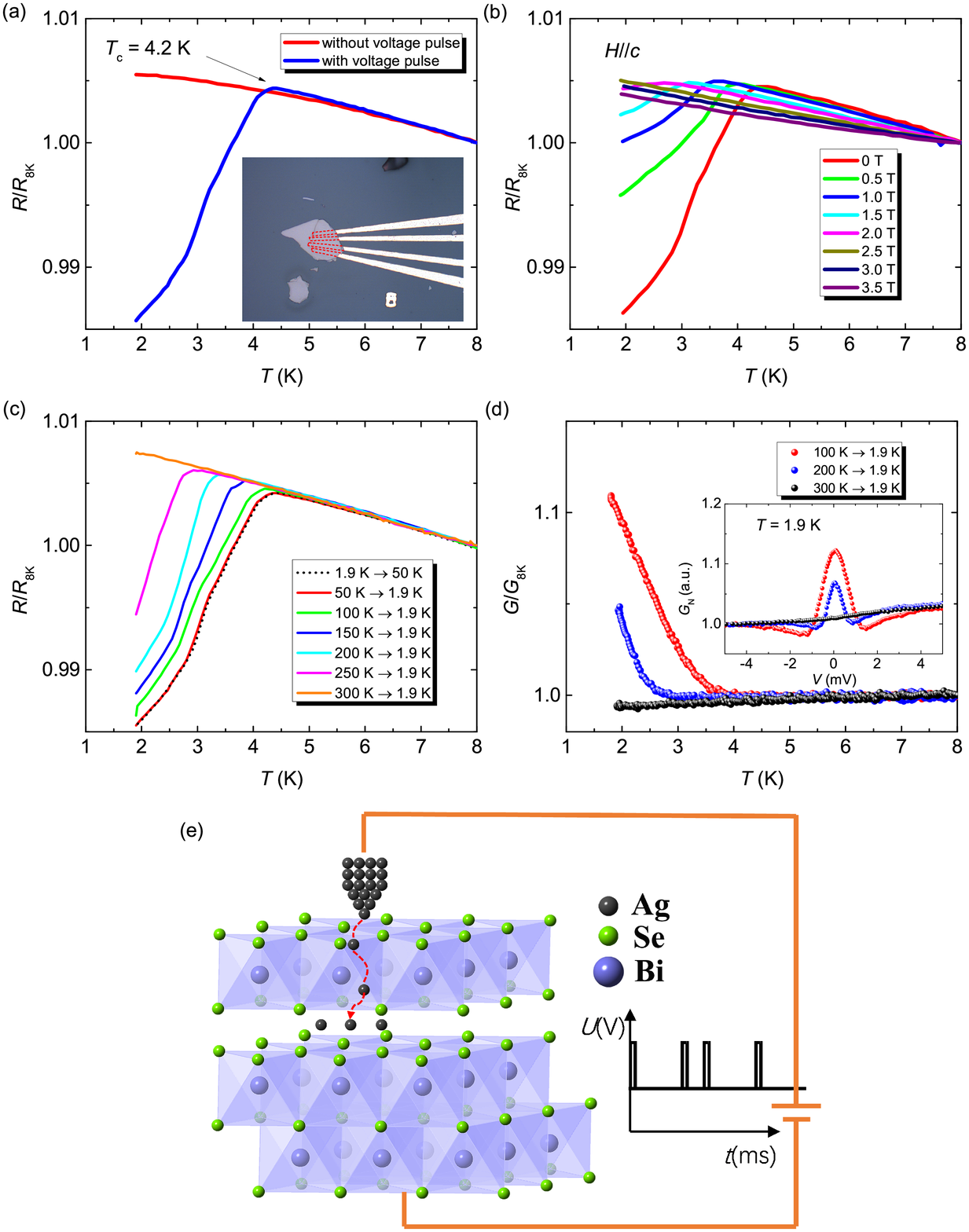}}
\vspace{-3pt} \caption{\label{Figure2} \textbf{Evidence of local and erasable superconductivity in Ag-Bi$_2$Se$_3$ contacts.} \textbf{(a)} Temperature dependence of the electrical resistance for an exfoliated Bi$_2$Se$_3$ flake before and after voltage pulse for comparison. The inset is the photo of an exfoliated flake with Ag electrodes pre-deposited on the bottom with the electrode distance $\sim$ 3 $\mu$m. \textbf{(b)} The evolution of R-T curves in different magnetic fields for the exfoliated Bi$_2$Se$_3$ sample after voltage pulse. \textbf{(c) \& (d)} Temperature dependence of the electrical resistance for exfoliated Bi$_2$Se$_3$, and ZBC for silver paint point-contacts after different thermal cycles with a temperature ramping speed of 3 K/min. The inset of \textbf{(d)} shows the Ag-Bi$_2$Se$_3$ point-contact conductance at 1.9 K after different thermal cycles. \textbf{(e)} Schematic illustration of Ag atoms locally trapped in the Bi$_2$Se$_3$ vdW gap in the contact area injected by voltage pulses.}
\vspace{-3pt}
\end{figure}

As for the underlying mechanism of the induced superconductivity in the Ag-Bi$_2$Se$_3$ type point-contacts, the voltage pulse can usually generate a considerable local electric field and inject the metallic atoms from the tip to Bi$_2$Se$_3$ \cite{PhysRevB.44.13703}. As shown in Fig. 2e, the injected Ag atoms should probably be trapped by and intercalated into the vdW gap of Bi$_2$Se$_3$ at low temperatures, forming some kind of Ag$_x$Bi$_2$Se$_3$ with superconductivity in the local region similar to the SC compound Cu$_x$Bi$_2$Se$_3$. We notice that SC in Ag$_x$Bi$_2$Se$_3$ has been theoretically proposed via the dynamic mean-field theory with a local density approximation and its $T\rm_c$ is estimated around 4.5 K, pretty similar to our experimental results \cite{koley2019superconductivity}. It is actually a common practice for atoms to be deposited on substrates in scanning probe microscopic lithography with the help of a voltage pulse on metallic tip \cite{PhysRevLett.65.2418, Saavedra_2010, garcia2014advanced}. In general, the atomic emission process is strongly dependent on the evaporation field for different atoms, where Ag has a relatively lower evaporation field in comparison with Cu and Au \cite{PhysRevB.44.13703, tsong1978field, katano2016creation}. On the other hand, even though Ti has a much lower evaporation field than Ag, the intercalation of Ti in Bi$_2$Se$_3$ probably would not induce SC as in Ag$_x$Bi$_2$Se$_3$ \cite{koley2019superconductivity}. Of course, the real atomic emission process can be more complicated and more detailed studies are further needed \cite{PhysRevB.44.13703, olsen2012surface}.

The disappearance of superconductivity in the Ag-Bi$_2$Se$_3$ mesoscopic point-contacts over thermal cycles probably originates from the instability of Ag intercalation in the Bi$_2$Se$_3$ vdW gaps at high temperatures. Considering the weak vdW gap between the Bi$_2$Se$_3$ quintuple layers, it is not surprising that no Ag$_x$Bi$_2$Se$_3$ compound has been reported to be synthesized at room temperatures so far. At 300 K, the intercalated Ag atoms would have escaped from the vdW gap thanks to the thermal activation energy. In comparison, even though SC in Cu$_x$Bi$_2$Se$_3$ crystals has been widely reported, its superconducting volume fraction shows a high dependence on quench conditions, where quenching from a lower temperature or not quenching at all can be detrimental to SC, strongly supporting the metastable nature of Cu intercalation \cite{PhysRevB.91.144506}. A metastable superconductivity by thermal cycles has also been reported in IrTe$_2$ with charge-density-wave (CDW) order as its ground state. For IrTe$_2$, a thermal quench process achieved by current pulses can induce or erase the SC in a tiny exfoliated sample, and it is a kinetic and nonequilibrium approach to induce SC as a metastable state competing with the CDW order \cite{oike2018kinetic, yoshida2018metastable}. In our case of Bi$_2$Se$_3$, the voltage pulse on point-contact can locally heat it to high temperatures and then rapidly cool down, mimicking the thermal quench process. However, the absence of SC in Au, Cu and Ti point-contacts on Bi$_2$Se$_3$ argues against the same SC mechanism as in IrTe$_2$, but favors the importance of Ag intercalation(Refer to Supporting Information for more discussions to exclude local strain, structural changes or impurity phase by voltage pulses as the the origin of superconductivity).

The erasable SC in mesoscopic point-contacts on Bi$_2$Se$_3$ with thermal cycles implies a promising application on logic and memory circuits as an electrical switch, if equipped with a local heater on the contact \cite{meijer2008wins, strukov2008missing, kwon2010atomic, szot2006switching}. A voltage pulse can momentarily drive the sample into superconducting state as a low-resistance phase, while an electrical heater can switch it back to the normal state as a high-resistance phase. This can serve as a phase-change memory while its different resistance states represent 1s and 0s for the stored digital data. The local and erasable superconductivity induced with Ag tip signifies a more accurate and controllable writing/design of superconducting circuits (even topological superconducting circuits) at low temperatures, if scanning probe microscopic lithography method can be introduced \cite{garcia2014advanced,tseng2005nanofabrication}. We notice a recent work carried out by the conductive-AFM method on two-dimensional electron gas and our results would facilitate such efforts \cite{PhysRevLett.120.147001, cen2008nanoscale}.

\begin{figure}
\centerline{\includegraphics[angle=0,width=0.99\textwidth]{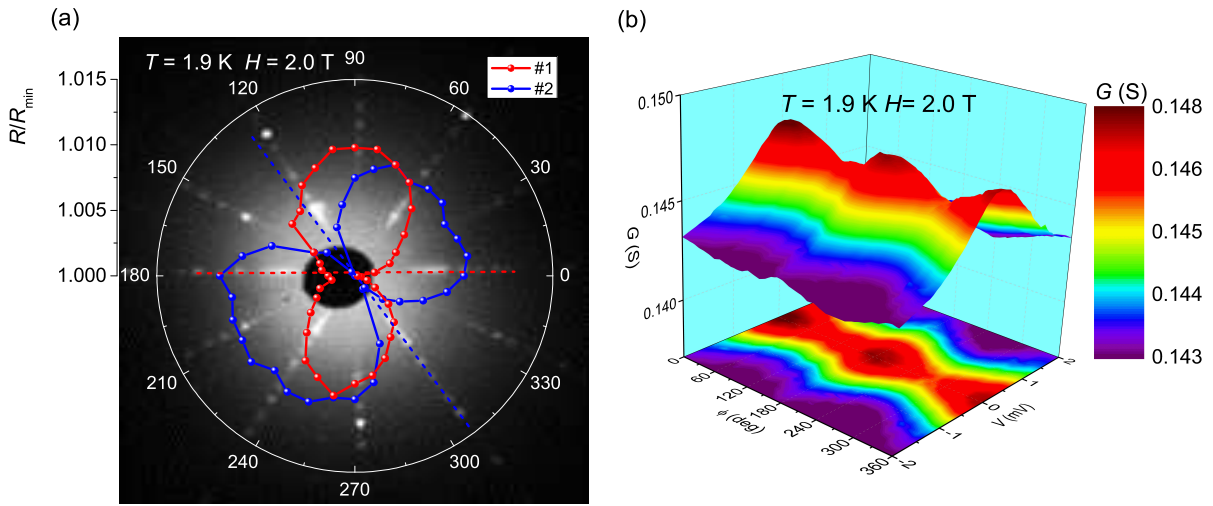}}
\vspace{-3pt} \caption{\label{Figure3}\textbf{Field-angle dependence of the soft-PCS of Ag-Bi$_2$Se$_3$ with the induced superconductivity.} \textbf{(a)} Field-angle dependence of the ZBR for two contacts \#1 and \#2 on the same Bi$_2$Se$_3$ crystal at 1.9 K with an in-plane field of 2.0 T. The red (blue) dashed line marks the short $C_2$ axis for its nematic domain \#1 (\#2), respectively. \textbf{(b)} Three-dimensional contour plot of the soft-PCS conductance curves as a function of field angle at 1.9 K and 2.0 T.}
\vspace{-3pt}
\end{figure}

Finally, we would like to investigate the superconducting nature of the Ag-doped topological insulator Bi$_2$Se$_3$ in our mesoscopic point-contacts, where our field-rotational PCS characterizes a rotational $C_4$ symmetry breaking and suggests an exotic topological SC as in Cu$_x$Bi$_2$Se$_3$ \cite{PhysRevB.90.100509, Matano2016, Yonezawa2017, PhysRevX.7.011009, PhysRevX.8.041024, PhysRevLett.123.027002, PhysRevB.94.180504}. The azimuthal field-angle dependence of the zero-bias point-contact resistance (ZBR) on Bi$_2$Se$_3$ after voltage pulse at 1.9 K and in field of 2.0 T is shown in Fig. 3a, where the magnetic field is rotated in the $ab$-plane. A clear 2-fold symmetry of the ZBR is present at 2.0 T, probably signaling a $C_2$ symmetry of the superconductor gap and thus a nematic SC despite of its hexagonal lattice structure. This 2-fold symmetry is only observed in the superconducting state below its upper critical field $H\rm_{c2}$ $\sim$ 3.0 T but absent in the normal state as shown in Fig. S2(a). When fixing the field at 2.0 T, the nematic behaviour disappears exactly around its local $T\rm_c$ as in Fig. S2(b) and Fig. S2(c). In our PCS configuration, the junction current is always along the $c$-axis and perpendicular to the rotational field, so that the influence of vortex motion and Lorentz effect can be excluded. In Fig. 3a, the azimuthal field-angle dependence of ZBR for two different point-contacts \#1 and \#2 on the same sample both shows an obvious dumbbell shape but with different long $C_2$ axis, which are perpendicular to the crystal axis and imply the existence of nematic SC domains in Bi$_2$Se$_3$ below $T\rm_c$. It is a strong evidence to exclude other artificial effects as the two-fold origins, such as sample misalignment, sample geometry or point-contact geometry. The existence of different nematic domains is probably caused by the local strain effect and its microscopic mechanism needs further careful studies.

Field-angle dependence of PCS conductance curves at 1.9 K and 2.0 T is shown in Fig. 3b as a three-dimensional contour plot, where a clear 2-fold symmetry can also be observed from both the peak intensity and width of Andreev reflection signals. The anisotropic magnitude of the nematic behaviour can be roughly estimated by the ratio of maximum/minimum conductance peak width with 1.5/1 $\sim$ 1.5, which is the result of anisotropic upper critical field $H\rm_{c2}$ in the $ab$-plane. In the Ginzburg-Landau theory, the $H\rm_{c2}$ is inversely proportional to the square-root product of coherence length $\xi$ in two separate directions orthogonal to the field. We can infer that the maximum of coherence length as well as the SC gap minimum in the $ab$-plane should be along the crystal axis. Such superconductivity is consistent with a fully gapped $\Delta \rm_{4y}$ state in an odd-parity $E\rm_u$ symmetry, implying a topological superconductor in the $D\rm_{3d}$ crystal point group \cite{PhysRevB.90.100509, PhysRevB.94.180504}. A similar ZBCP in the PCS on Cu$_x$Bi$_2$Se$_3$ has been claimed due to Majorana fermions \cite{PhysRevB.90.100509, PhysRevLett.107.217001}, however, we note it can also arise from thermal effect for point-contacts in thermal regimes \cite{PhysRevB.69.134507, 0953-2048-23-4-043001}. Further studies are required to explore the nature and origin of ZBCP for the induced SC in our Ag-Bi$_2$Se$_3$ point-contacts.

In conclusion, we have discovered the unexpected superconductivity only in the Ag-Bi$_2$Se$_3$ mesoscopic point-contacts by applying a voltage pulse, either in soft- or mechanical- PCS setup, which is rather absent in the case of Au, Cu or Ti contacts within the voltage limit. We propose that the voltage pulse and thus electrical field should inject Ag atoms into the vdW gap of Bi$_2$Se$_3$ and induce a local superconductivity. The superconductivity can be erased by thermal cycles, where warming samples up to high temperatures seems to lose the trapped Ag atoms from the weak vdW gap. Our discovery of the erasable and repetitive superconductivity in topological insulator Bi$_2$Se$_3$ may pave a new route for applications of quantum electronic devices that harness the power of topological superconductivity and Majorana fermions.

\textbf{Materials}

The high quality Bi$_2$Se$_3$ single crystals were offered by Prmat (Shanghai) Technology Co., Ltd..

\textbf{Resistivity measurement}

Electrical resistivity of Bi$_2$Se$_3$ was measured by the conventional four-probe method. Electrodes on bulk sample of a larger size were made with silver paint (SPI05001-AB), which can be dry within several minutes, while those on the exfoliated Bi$_2$Se$_3$ flake were made with pre-sputtered Ag electrodes on Si/SiO$_2$ chip and the Bi$_2$Se$_3$ flake was  mechanically transferred on the top of Ag electrodes.

\textbf{Point-contact spectroscopy measurements}

Soft point-contacts on Bi$_2$Se$_3$ were prepared by attaching a 20 $\mu$m diam. gold wire with a silver-paint drop at the end on the freshly-cleaved surface at room temperature. In such a configuration, thousands of parallel nanoscale channels were assumed between individual silver particles and the crystal surface. Mechanical point-contacts on Bi$_2$Se$_3$ in a needle-anvil style were prepared by engaging a sharp tip on the sample surface by piezo-controlled nano-positioners. The tips can be Ag, Au, Cu and Ti wires, with sharp apex cut by razor blade. The conductance curves as a function of bias voltage, $G$($V$), were recorded with the conventional lock-in technique in a quasi-four-probe configuration. The output current is mixed with dc and ac components, which were supplied by the model 6221 Keithley current source and model 7265 DSP lock-in amplifier, respectively. The first harmonic response of the lock-in amplifier is proportional to its point-contact resistance d$V$/d$I$ as a function of the biased-voltage $V$.

\textbf{Low temperature measurements}

Resistivity and point-contact spectroscopy down to 1.9 K were measured in a Quantum Design Physical Property Measurement System (14T-PPMS) equipped with a sample rotator.

\textbf{Pulse application}

The voltage pulse is applied by signal generator(Rigol DG 1022) with pulse waveform.

\textbf{Acknowledgements} \par 
We are grateful for valuable discussions with Z.A. Xu, H.Q. Yuan, C. Cao and Y. Liu. This work has been supported by the National Key Research \& Development Program of China (Grant No. 2016YFA0300402, No. 2017YFA0303101 and No. 2018YFC0810200) and the National Natural Science Foundation of China (Grant No. 11674279, No. 11374257 and No. 21703204). X.L. would like to acknowledge support from the Zhejiang Provincial Natural Science Foundation of China (LR18A04001). X.W. would like to acknowledge support from Key Research and Development Project of Zhejiang (No. 2018C01048), Zhejiang Lab (No.2018EB0ZX01). We also thank Prmat (Shanghai) Technology Co., Ltd. for their offered Bi$_2$Se$_3$ single crystals.

\end{document}